\documentclass[usenatbib,preprint,dvipdfmx]{mn2e}
\usepackage{graphicx}
\usepackage {natbib,aas_macros}
\usepackage {bm}
\usepackage {amsmath,amssymb}
\usepackage {setspace}
\newcommand{\ltsim}{\protect\raisebox{-0.5ex}{$\:\stackrel{\textstyle <}{\sim}\:$}}
\newcommand{\gtsim}{\protect\raisebox{-0.5ex}{$\:\stackrel{\textstyle >}{\sim}\:$}}
\newcommand{\cs}{c_{\rm s}}

\usepackage{natbib, aas_macros}
\title[Rotation Profiles in Star Forming Clouds]{The Origin of the Rotation Profiles in Star Forming Clouds}

\author[S. Z. Takahashi et al.]{S. Z. Takahashi$^{1,2}$, K. Tomida$^{3,4}$, M. N. Machida$^{5}$, and S. Inutsuka$^{6}$\\
$^1$Astronomical Institute, Tohoku University,6-3 Aoba, Aramaki-aza, Aoba-ku, Sendai, Miyagi, Japan\\%{sanemichi@astr.tohoku.ac.jp}
$^2$Department of Physics, Kyoto University, Oiwake-cho, Kitashirakawa, Sakyo-ku, Kyoto 606-8502, Japan\\
$^3$Department of Earth and Space Science, Osaka University, Machikaneyama-cho, Toyonaka-shi, Osaka 560-0043, Japan\\%{tomida@vega.ess.sci.osaka-u.ac.jp}
$^4$Department of Astrophysical Sciences, Princeton University, Princeton, NJ 08544, USA\\
$^5$Department of Earth and Planetary Science, Kyushu
University, Higashi-ku, Fukuoka 812-8581, Japan\\%machida.masahiro.018@m.kyushu-u.ac.jp
$^6$Department of Physics, Nagoya University, Furo-cho, Chikusa-ku, Nagoya, Aichi 464-8602, Japan%{inutsuka@nagoya-u.jp}
}

\begin{document}
\maketitle
\begin{abstract}
 Angular momentum distribution and its redistribution are of crucial importance in the formation and evolution of circumstellar disks. 
Many molecular line observations toward young stellar objects indicate that radial distributions of the specific angular momentum $j$ 
have a characteristic profile.
In the inner region, typically $R\ltsim 100 \, {\rm AU}$, the specific angular momenta distribute like $j\propto r^{1/2}$, indicating existence of rotationally supported disk. 
In outer regions, $R\gtsim 5000 \, {\rm AU}$, $j$ increases as the radius increases and the slope is steeper than unity, which is supposed to reflect the original angular momentum distributions in the maternal molecular clouds. 
In the intermediate region, $100\,  {\rm AU} \ltsim R \ltsim 5000 {\rm AU}$, $j$-distribution appears almost flat.
While this is often interpreted as a consequence of conservation of the specific angular momentum, the interpretation actually is insufficient and requires a stronger condition that the initial distribution of $j$ must be spatially uniform. 
However, this requirement seems to be unrealistic and inconsistent with observations. 
In this work, we propose a simple alternative explanation; the apparently flat $j$ profile is produced by strong elongation due to the large velocity gradient in the accreting flow no matter what the initial $j$ distribution is. 
To show this we provide a simple analytic model for gravitational collapse of molecular clouds. 
We also propose a method to estimate ages of protostars using solely the observed rotation profile.
We demonstrate its validity in comparison with hydrodynamic simulations, and apply the model to young stellar objects such as L1527 IRS, TMC-1A and B335.
\end{abstract}


\section{Introduction}
Protoplanetary disks are formed as natural by-products of star formation processes. While extensive observations and theoretical studies have been carried out, formation and evolution of the disks still remain the subject of active investigation. 
Obviously, the most important factor here is distribution and transport of the angular momentum in star forming clouds. In particular, distribution of the angular momentum in infalling envelopes is very important because it will control future evolution of circumstellar disks, and also because it can retain some information about the initial conditions of star formation processes.

Extensive observations have been performed to retrieve the dynamical information within star forming clouds. 
Various molecular lines are used depending on densities and scales of interest. 
Compiling those results, apparently there is a universal trend in the rotation profiles. 
\cite{ohashi97} have found the relations between the trend of the rotation profiles and dynamics of the gravitational collapse of molecular cloud cores.
In the large scale, typically $R\gtsim 5000\, {\rm AU}$, the specific angular momentum $\mathbf{j}\equiv \mathbf{r}\times \mathbf{v}$ increases as the radius increases like $j \propto R^{1.6}$ \citep[e.g. ][]{goodman93}. 
This distribution likely reflects the initial conditions of the molecular cloud cores, possibly the turbulent velocity fields following the Larson's law \citep{lrs81}. 
On the other hand, rotating protoplanetary disks with Keplerian rotation have been observed in the small scales ($R\ltsim 100\, {\rm AU}$) around the central young stellar objects \citep[e.g.][]{tobin12}.
While the envelopes still maintain some information about the initial conditions, the disks with the  Keplerian rotation have experienced angular momentum transport and forgotten the initial conditions.
In the intermediate scales between the Keplerian disks and molecular clouds, where the gas in the envelope is infalling, interestingly the specific angular momentum distribution appears almost flat ($j \propto R^0$, Fig. 6 of \cite{ohashi97} ). 
\cite{ohashi97} have pointed out that this can be interpreted as a consequence of ``conservation of angular momentum'' in the infalling envelope \cite[see also][]{2013EAS....62...25B,2014prpl.conf..173L}. 
In this scale, the specific angular momentum is indeed conserved because the infall speed is much larger than the rotation speed, or in other words, the dynamical (free-fall) time scale is much shorter than that of angular momentum transport. 
Although the conservation of the angular momentum itself is physically plausible, it is not sufficient to explain the flat distribution because the conservation of angular momentum only means that the specific angular momentum of a gas element is conserved in time, i.e. $\frac{dj}{dt}=0$. 
On the other hand, the flat distribution of specific angular momentum means that the specific angular momentum of each gas element in the intermediate region is almost the same.
In order to make the angular momentum distribution spatially uniform, this argument additionally requires that the initial specific angular momentum distribution must be uniform; $\left.\frac{dj}{dR}\right|_{t=0}=0$ or $j = Rv_\phi$ is constant. This is, however, quite unlikely as it is contradictory to observations indicating larger velocity dispersions in larger scales.
Moreover, this flat profile is virtually universal in many observed star forming clouds. 
Therefore, we need a robust physical mechanism to explain this appearance, which is independent from or only weakly depends on specific initial conditions.

Gravitational collapse of rotating clouds is investigated by numerous previous studies. 
\citet{1981Icar...48..353C} investigated the formation and evolution of circumstellar disks approximating the gravitational collapse of molecular clouds as spherical collapse in an outer region and ballistic trajectories in an inner region.
\citet{1984ApJ...286..529T} extended the self-similar solution of the inside-out collapse of a singular isothermal sphere \citep{1977ApJ...214..488S} to include the effect of rotation. Numerical simulations of gravitational collapse of rotating clouds are also investigated and self-similar behaviors are found \citep{1984PThPh..72.1118N,1997ApJ...478..569M}.
\citet{1998ApJ...493..342S} obtained a self-similar solution of gravitational collapse of rotating a thin disk and discussed the evolution of the disks before and after the protostar formation.
\citet{2011ApJ...742...57Y} and \cite{2013ApJ...772...22Y} compared the analytic models of gravitational collapse of rotating clouds with the observed angular velocity distributions
These studies adopted isothermal spheres with rigid-body rotation as initial conditions and assumed that the time evolution of the clouds were given by the inside-out collapse.

In this work, we develop an analytic model of unmagnetized infalling envelope that reproduces the results of numerical simulations in order to investigate the time evolution of infalling envelope. 
This model does not use the singular isothermal sphere but can be used for general spherically symmetric initial density profile. 
In this paper, we discuss the origin of these rotational profiles in star forming clouds. 
In particular, we show that while the angular momentum is conserved at the intermediate scale, the flat distribution is a consequence of elongation due to the velocity gradient in infalling envelopes. 
This model can be used for estimating the ages of protostars solely from the rotation profiles, and we apply it to well-studied young stellar objects.
This paper is organized as follows. In \S2 we construct an analytic model of the angular momentum distribution in collapsing clouds. We demonstrate its validity in comparison with numerical simulations in \S3 and provide case studies for well-studied young stellar objects with circumstellar disks such as L1527 IRS, TMC-1A and B335 in \S4. \S5 and \S6 are devoted to discussion and conclusions.

\section{Analytic Model}
\label{analytic_model}

In this section, we construct an analytic model of infalling envelopes. The goal is to provide a simple but useful model explaining the angular momentum distributions in observations and numerical simulations.

Before the gas element gets close to its centrifugal radius, it falls almost radially. Therefore, the effect of rotation is not dynamically important and is negligible in the outer region
\footnote{ It is straightforward to include the effect of centrifugal force in our calculation.  Using an analytic model with centrifugal force, we test how much the infalling motion of the envelope is affected by the centrifugal force. We confirm that it becomes important only in the inner region near the centrifugal radius (typically $r<100\,$AU).}
\cite[cf.][]{1981Icar...48..353C}. 
In reality, highly non-linear angular momentum transport by self-gravity or magnetic fields dominates the evolution in that small-scale, which anyway is very difficult to handle analytically. Thus we only discuss the infalling envelope here. We exclude the region near the central object from our model assuming that a circumstellar disk with a Keplerian rotation profile is formed there.

In this situation, we can assume that the accretion flow is spherically symmetric; the gravitational force is approximated to be $\sim \frac{GM_r}{r^2}$ where $M_r$ is the enclosed mass within the radius $r$, and the specific angular momentum $\mathbf{j}\equiv \mathbf{r}\times \mathbf{v}$ is conserved and passively advected. Note that $M_r$ is constant on the gas element's frame in this treatment; $M_{r(t)}=M_{r(t=0)}$, so hereafter we omit the subscript. Also, we assume that the gas collapses isothermally, i.e. $p=\rho c_s^2$ where $c_s$ is the sound speed and is constant. Then, the equation of motion of the gas element for the radial velocity $v_r$ becomes: 
\begin{eqnarray}
\frac{dv_r}{dt}&=&-\frac{1}{\rho}\frac{\partial p}{\partial r}-\frac{GM}{r^2}\nonumber\\
&=&-\frac{p}{r\rho}\frac{\partial \ln p}{\partial \ln r}-\frac{GM}{r^2}.
\end{eqnarray}
Here, $\frac{\partial \ln \rho}{\partial \ln r}$ is order of unity; for example, it is $-2$ when the density profile is close to the Larson-Penston solution, $\rho\propto r^{-2}$ in the outer region \citep{lrs69,pen69}. We approximate this value is constant in time but depends on the initial radius of the gas element; $\frac{\partial \ln \rho}{\partial \ln r} = \frac{\partial \ln \rho_{\rm i}}{\partial \ln r_{\rm i}} \equiv \beta (r_i)$. Under this assumption, the pressure force is proportional to $r^{-1}$; $\frac{\partial p}{\partial r}\sim \beta p/r$. 
Since the gravitational force is proportional to $r^{-2}$, the ratio of the  pressure force to the gravitational force decreases as the radius of the gas element decreases. In other words, the gas pressure becomes less important as the gravitational collapse proceeds. Therefore, we estimate $\beta(r_{\rm i})$ by using initial density distribution to reproduce the initial pressure force \cite[]{2013ApJ...770...71T}. Because this gas pressure decreases the infall velocity, the infall velocity is smaller than the free-fall velocity. Using $\beta(r_{\rm i})$ estimated from initial density distribution, we can calculate the motion of the gas element locally including the effect of the gas pressure without knowing the density distribution, and greatly simplify the equation of motion: 
\begin{figure*}
\begin{center}
\scalebox{1}{\includegraphics{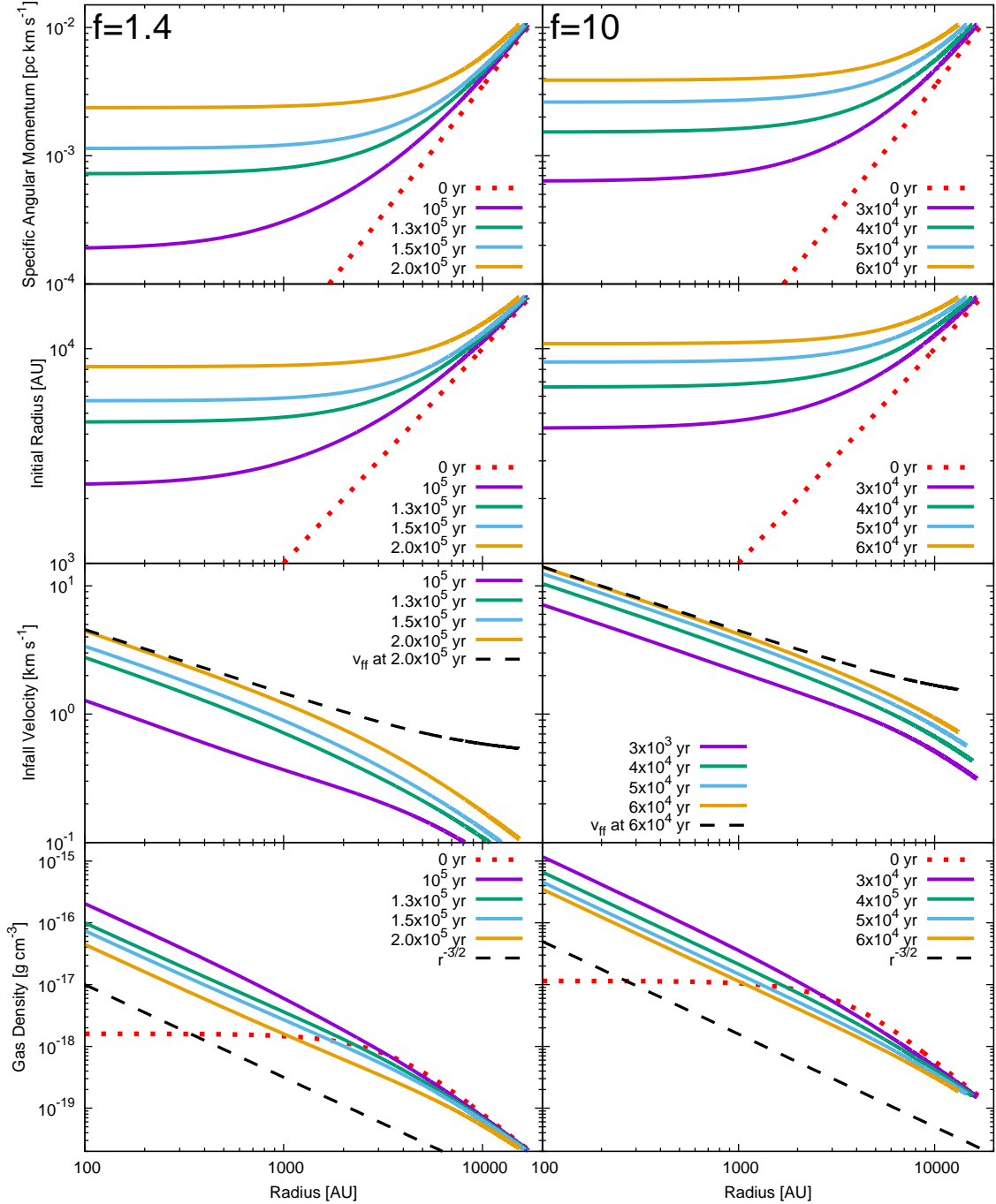}}
\caption{The radial profiles of the specific angular momentum in the mid-plane, initial radius, infall velocity and gas density from top to bottom. Only $R > 100 $ AU is plotted because of the limitation of the model. In the figure of the infall velocity, we also plot the free-fall velocity $v_{\rm ff} = \sqrt{2GM_{\rm r}/r}$ at $t= 2\times 10^5$ yr. We adopt $f=1.4$ and $f=10$ for the left and right panels, respectively. Red dotted lines show the initial profiles of specific angular momentum, initial radius and gas density. The initial density profiles are given by the Bonnor-Ebert sphere density profile increased by factor of $f$. }
\label{ana}
\end{center}
\end{figure*}
\begin{eqnarray}
\frac{dv_r}{dt}&=&-\frac{\beta c_s^2}{r}-\frac{GM}{r^2}.
\end{eqnarray}
Then we introduce another non-dimensional value $f$ which is proportional to the ratio between the gravitational potential to the thermal energy density at the initial radius; $f (r_i)\equiv -\frac{GM}{r_{i}\beta c_s^2}$. Note that $f(r_i)$ is usually positive because $\beta$ is typically negative in a collapsing cloud. Also, $f(r_i)$ is constant in time but spatially not uniform in general, and it is determined by the initial condition.
\begin{eqnarray}
\frac{dv_r}{dt}&=&\frac{GM}{f r_ir}-\frac{GM}{r^2}\nonumber\\
&=&GM\left(\frac{1}{f r_i r}-\frac{1}{r^2}\right).\label{eq:dvrdt}
\end{eqnarray}
Integrating this equation after multiplying $v_r \equiv \frac{dr}{dt}$ yields:
\begin{eqnarray}
\frac{1}{2}\left(\frac{dr}{dt}\right)^2&=&GM\int_{r_i}^r\left(\frac{1}{f r_ir}-\frac{1}{r^2}\right)dr\nonumber\\
&=&GM\left[\frac{1}{f r_i}\ln\left(\frac{r}{r_i}\right)+\frac{1}{r}-\frac{1}{r_i}\right].\\\nonumber\\
\frac{dr}{dt}&=&-\left[2GM\left\{\frac{1}{f r_i}\ln\left(\frac{r}{r_i}\right)+\frac{1}{r}-\frac{1}{r_i}\right\}\right]^{\frac{1}{2}}.
\end{eqnarray}\\
Here we have taken the negative sign because we are interested in a collapsing solution. We define a new dimensionless radius $x\equiv \frac{r}{r_i}$ and then integrate this equation.
\begin{eqnarray}
t&=&-\sqrt{\frac{r_i^3}{2GM}}\int_1^x\frac{dx}{\sqrt{f^{-1}\ln x+x^{-1}-1}}\nonumber\\
&=&\sqrt{\frac{3}{8\pi G \bar{\rho}_{r_i}}}\int_x^1\frac{dx}{\sqrt{f^{-1}\ln x+x^{-1}-1}} \nonumber,\\
&=&\frac{2}{\pi}t_{\rm ff}\int_x^1\frac{dx}{\sqrt{f^{-1}\ln x+x^{-1}-1}} \label{pos},
\end{eqnarray}
where $\bar{\rho}_{r_i}=\frac{3M_{r_i}}{4\pi r_i^3}$ is the mean gas density within $r_i$ and $t_{\rm ff}=\frac{\pi}{2}\sqrt{\frac{r_i^3}{2GM}}$ is the free-fall time. This equation gives the location of a gas element, $r = xr_{\rm ini}$, as a function of time and initial radius. 
Using this equation, we can calculate distribution of the specific angular momentum given an initial condition. Because $j$ is conserved, it is simply transported by the motion of the gas element; $j(r(t),t)=j(r(0),0)$.

For later use, we calculate how long it takes for a gas element at an initial radius $r_i$ to reach the center of the cloud. This time $t_{\rm c}$ can be calculated from Equation (\ref{pos}) by taking the limit of $x\rightarrow 0$. This time depends only on $f$ if it is normalized by the free-fall time $t_{\rm ff}$.
\begin{eqnarray}
\frac{t_{\rm c}}{t_{\rm ff}}(f)=\frac{2}{\pi}\int_0^1\frac{dx}{\sqrt{f^{-1}\ln x+x^{-1}-1}}.\label{tc}
\end{eqnarray}
When $f$ is large and the pressure force is not significant, the right hand side of Equation (\ref{tc}) is unity, and the time that a gas element takes to reach the center is given by the free-fall time.

Evolution of the density distribution can be also calculated from Equation~(\ref{pos}). Since the mass within a shell of thickness $dr$, $dM= 4\pi \rho r^2 dr$, is constant, we can calculate the density distribution from the following equation:
\begin{eqnarray}
\rho=\rho_{\rm ini}\left(\frac{r_{\rm ini}}{r}\right)^2\frac{dr_{\rm ini}}{dr}.\label{eq:rho}
\end{eqnarray}
$r$ and $dr$ are obtained from Equation~(\ref{pos}).
Using Equation (\ref{eq:rho}) and the $r$ derivative of Equation (\ref{pos}), we obtain
\begin{eqnarray}
 \rho(x) = \left[\frac{3}{2}
	\left(\frac{1}{ \bar{\rho}_{r_i}} - \frac{1}{\rho_{\rm ini}}
	\right)
	x^2\sqrt{f^{-1}\ln x + x^{-1} -1} \right.\nonumber \\
	\times\left.\int_1^x\frac{dx}{\sqrt{f^{-1}\ln x + x^{-1} -1}}
	+ \frac{x^3}{\rho_{\rm ini}}\right]^{-1},
\end{eqnarray}
where we assume that $f$ is spatially uniform.
Hereafter we use a model that $f$ is spatially constant, which is explained as follows.

For example, let us consider a super-critical Bonnor-Ebert \citep[][hereafter BE]{bonnor,ebert} sphere with solid-body rotation as the initial condition. This initial condition is often used in numerical simulations to model an isolated dense molecular cloud core. First we construct a critical BE sphere with a central density of $n_c = 3 \times 10^5 \, {\rm cm^{-3}}$, a temperature of $T = 10$ K, a radius of $R=17,400 \, {\rm AU}$ and an angular velocity of $\Omega = 4.8\times 10^{-14} \, {\rm s^{-1}}$. Because the BE sphere is an equilibrium state, it does not collapse immediately. To make it gravitationally unstable, we increase the  whole density of BE sphere by $40\%$ ($f=1.4$). This method has been often used in calculations for the gravitational collapse of cloud cores \cite[e.g.][]{1977ApJ...218..834H, 2010ApJ...724.1006M}
 The evolution of the profiles in the disk mid-plane (the plane that perpendicular to the rotation axis, where the specific angular momentum of the gas element is $r_{\rm i}^2 \Omega$) are shown in the left panels of Figure~\ref{ana}. 
Because we consider that the angular momentum is transported purely passively, this model cannot account for a rotationally supported disk and is valid only in the region beyond the centrifugal radius. 
To show the dependence of the analytic model on the parameter $f$, the evolution of the radial profiles of the specific angular momentum in the mid-plane, initial radius, infall velocity and gas density with $f=10$ are shown in the right panels of Figure~\ref{ana}. 
The other parameters are the same as those used for the previous calculation whose results are shown in left panels of Fig.~\ref{ana}. The nature of all profiles is the same as the nature of the profiles shown in left panels in Fig.~\ref{ana}, except the timescale of the gravitational collapse. 
For larger $f$, the collapse timescale becomes shorter because the gas pressure is less significant and the free-fall times is shorter due to the higher initial density.

The model successfully reproduces the tendency observed in star forming clouds. 
While the outer region of the infalling envelope maintains the initial angular momentum distribution, the specific angular momentum looks almost constant in the inner region. 
While the specific angular momentum is indeed conserved in this model, what actually is crucial is the radial velocity gradient in the infalling envelope. 
The gas element originally located at a smaller radius attains higher radial velocity than one at a greater radius due to the stronger gravity. 
This property is common in any run-away collapse solution including the Larson-Penston solution. 
This radial velocity gradient elongates the infalling envelope radially.
It is clearly visible in the plot of the initial radius of Fig \ref{ana} that the gas elements in the inner region originally come from similar initial radii. 
Because the specific angular momentum of gas elements is conserved,
the elongation of the infalling envelope elongates the initial angular distribution and makes it look flat. 
Therefore the region with constant specific angular momentum can be formed simply as a consequence of strong elongation in a run-away collapse of a cloud, regardless of the initial angular momentum distribution.
This flat specific angular momentum distribution also appears in the self-similar solutions of the gravitational collapse of rotating spherical clouds \citep{1984ApJ...286..529T} and  flattened disks \citep{1998ApJ...493..342S}.
The flat region elongates and the specific angular momentum in this region gets higher as time passes by, which can be used as an indicator of the age of the system. The advantage of this method is that it only uses the kinetic information which can be easily acquired, and does not rely on the stellar evolution model. We demonstrate this in Section \ref{comparison} in comparison with observations.

\section{Numerical Simulation}
\label{simulation}
\begin{figure*}
\begin{center}
\includegraphics[width=16cm]{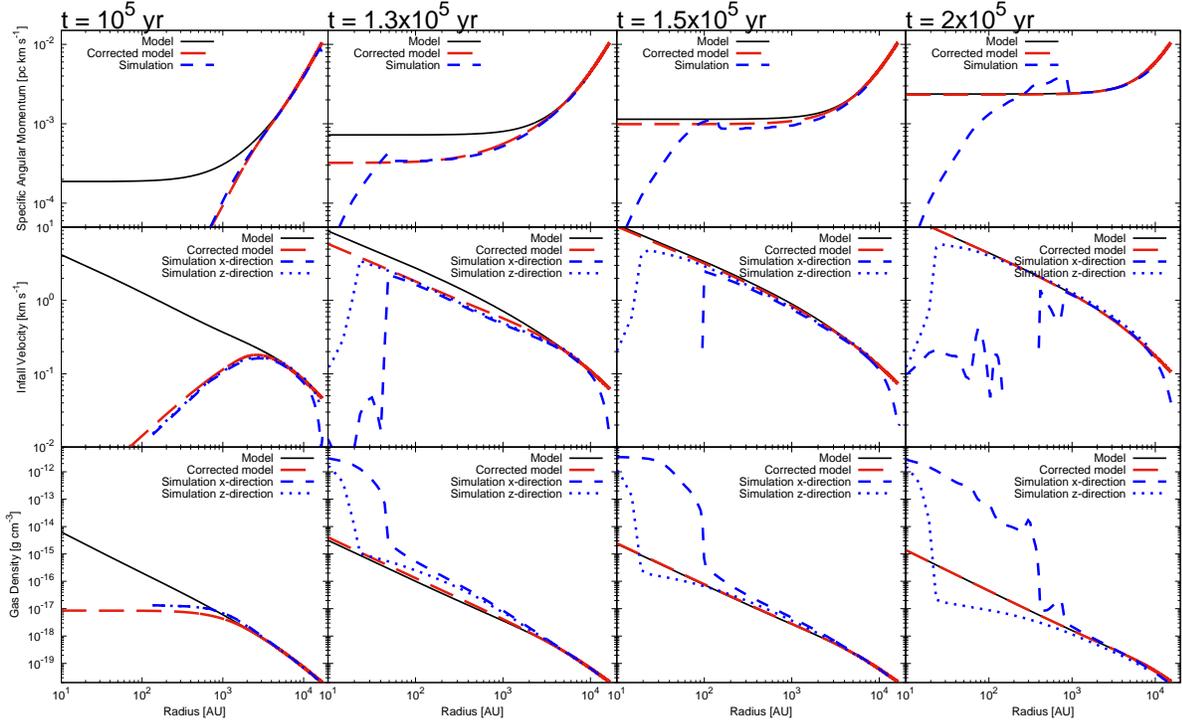}
\caption{Comparison between the hydrodynamic simulation and the analytic solution at $t=10^5$ yr  $1.3 \times 10^5$ yr,  $1.5 \times 10^5$ yr and $ 2.0 \times 10^5 $ yr(from left to right), the specific angular momentum, infall velocity and gas density from top to bottom. The solid lines are the analytic models. The distributions in the disk mid-plane in simulation are plotted with the dashed lines, while the dotted lines show the distributions along the rotational ($z$) axis. The red dashed lines show the analytic models with a correction factor. We adopt f=1.4 for both the hydrodynamic simulation and the analytic solution.}
\label{sim}
\end{center}
\end{figure*}

In order to demonstrate the validity and limitation of the analytic model, we compare it with a hydrodynamic simulation. We perform a three-dimensional nested-grid hydrodynamic simulation with self-gravity \citep{mcd05b,mcd05a,tomida13,2015ApJ...801..117T}. For hydrodynamic part, the second-order accurate MUSCL scheme and Roe's Riemann solver are used. Self-gravity is solved with the Multi-grid solver developed by \citet{mh03}. 
The basic equations are as follows:
\begin{eqnarray}
\frac{\partial \rho}{\partial t} + \nabla \cdot (\rho \bm{ v})&=&0,\\
\rho\frac{\partial \bm{v}}{\partial t} + \rho(\bm{v} \cdot \nabla ) \bm{v} &=& -\nabla P -\rho\nabla \phi. 
\end{eqnarray}
The gravitational potential is composed of two parts:
\begin{eqnarray}
\phi &= &\phi_{\rm gas} + \phi_{\rm ps},\\
\nabla^2 \phi_{\rm gas}&=&4\pi G \rho,\\
\phi_{\rm ps}&=&-\frac{GM_{\rm sink}}{r},
\end{eqnarray}
where $M_{\rm sink}$ is the mass of the sink particle described below, and $\phi_{\rm ps}$ and $\phi_{\rm gas}$ are the gravitational potential due to the sink particle and the gas.
In order to mimic the thermal evolution, the following barotropic approximation is used: 
\begin{eqnarray}
T={\rm max}\left(10, 10\times \left(\frac{\rho}{\rho_{cr}}\right)^{\gamma-1} \right),
\end{eqnarray}
where $\rho_{cr} = 1.0\times 10^{-13}\, {\rm g \, cm^{-3}}$ is the critical density that the gas becomes adiabatic, and the effective adiabatic index $\gamma$ is $5/3$. 
In order to follow the long-term evolution, we introduce a sink particle. 
The calculation is started without the sink particle. 
The sink particle is placed at the centre of the cloud core when the density within radius $r_{\rm sink}=$1 AU becomes greater than the threshold density  $\rho_{\rm sink}=3.8\times 10^{-12}\, {\rm g\, cm^{-3}}$. The gas exceeding the threshold density
in the region $r<r_{\rm sink}$ is removed, and the mass of the removed gas added to the mass of the sink particle $M_{\rm sink}$. 
Note that, however, the details of the simulation are not of crucial importance because we only discuss the structure of the infalling envelope.

The initial condition is the same as in the previous section, a super-critical BE-like sphere (the central density is $n_c = 3 \times 10^5 \, {\rm cm^{-3}}$, a temperature is $T = 10$ K, a radius is $R=17,400 \, {\rm AU}$, an angular velocity of $\Omega = 4.8\times 10^{-14} \, {\rm s^{-1}}$, and the density is increased by $40\%$ ($f=1.4$)). 
All the variables at the surface of the BE sphere is fixed to the initial values to model a dense molecular cloud core confined in a static interstellar medium. 
The results at $t= 10^5$ yr, $1.3 \times 10^5$ yr, $ 1.5\times 10^5$ yr, and $2.0\times 10^5$ yr are shown in Figure~\ref{sim} along with the analytic solution. 
The blue dotted and dashed lines show the results of the numerical simulation and thin black lines show the results of the analytic model.
At $t=10^5 $ yr, the analytic model and the numerical simulation show the different distributions.
This is because the protostar is not formed in the numerical simulation, although it is already formed in the analytic model.
The density and velocity distributions before and after the protostar formation are completely different: the former are similar to the Larson-Penston solution \cite[]{lrs69, pen69} and the latter are similar to the Shu solution \cite[]{1977ApJ...214..488S}.
This result suggests that the analytic model tends to underestimate the infall time of the envelope (Equation (\ref{pos})) and the protostar formation time.
The underestimation of the infall time comes from the modeling of the pressure.
In our model, the ratio of pressure gradient to gravitational forces decreases monotonically as the gas collapses toward the center.
However, the numerical simulation shows that the ratio increases with time at first in the inner region where the included mass is smaller than about the Jeans mass. 
This means that the analytic model underestimates the pressure gradient force acting on the inner region, and this results in the underestimation of the infall time.

To avoid the underestimation of the infall time before the protostar formation, we introduce a correction factor into Equation (\ref{pos}). 
The corrected infall time is given as follows,
\begin{equation}
 t=A(M)\sqrt{\frac{r_i^3}{2GM}}\int_x^1\frac{dx}{\sqrt{f^{-1}\ln x+x^{-1}-1}},
\label{eq:correctedtime}
\end{equation}
\begin{equation}
 A(M) = 1+0.4\exp\left(-\frac{2M}{M_{\rm J}}\right),
\label{eq:correctionfactor}
\end{equation}
where $M_{\rm J}$ is the Jeans mass defined by the central density.
The correction factor $A(M)$ increases the infall time for $M \lesssim M_{\rm J}$.
The results of the corrected models are shown by the red dashed lines in Figure \ref{sim}.
The corrected model can reproduce the numerical simulation at $t=10^5$ yr. The results of the model at the other times also reproduce the simulation better than that of the previous model. 
We use the model with the correction factor $A(M)$ in the following sections.
Note that our model can reproduce the pressure gradient force for the outer region where the included mass is larger than about the Jeans mass.
Thus, for the later stage of the collapse, the results of the model without the correction factor are similar to those of the numerical simulation formation.
Even at $t = 1.3 \times 10^5 $ yr, the specific angular momentum and infall velocity of the numerical simulation and the analytic model without the correction factor agree with each other well, within a factor of 2.2.

In the simulation, Rotationally-supported circumstellar disks with a radius of $R\sim 50$ AU at $t=1.3\times 10^5 $ yr, $R\sim 100$ AU at $t=1.5\times 10^5$ yr, and $R\sim 900$ AU at $t=2.0 \times 10^5$ yr are formed. 
The analytic solution is not applicable to this region because the effect of the centrifugal force is not included in the model. 
The analytic solution reproduces the distribution along the rotational axis ($z$, dotted lines) better because the angular momentum is passively transported and has no dynamical effect in the analytic model. 

\section{Estimates of Protostar Ages}
\label{comparison}
In this section, we discuss how to apply our model to real observations of circumstellar disks around young stellar objects. In particular, we demonstrate that our model can provide an independent estimate of the age of the system. 
Hereafter, we focus on the dynamics of infalling envelope in the mid-plane.

In order to apply our theoretical model, we first need to estimate
the initial density distribution. Because we cannot observe the
initial conditions, we assume a super-critical BE-like sphere as in
Section~2, with parameters $f = 1.4$ and $T = 10$ K.
The uncertainty associated with $f$ is discussed later in Section \ref{discussion}. The initial density distribution, which is proportional to that of BE sphere, is approximately given by
\begin{equation}
 \rho(r) = \rho_{\rm c}\left(1+\frac{\xi^2}{\xi_{\rm c}^2}\right)^{-3/2},\label{eq:rho_r}
\end{equation}
where $\rho_{\rm c}$ is the central density, $\xi$ is a normalized radial coordinate,
\begin{equation}
 \xi = \left(\frac{4\pi G \rho_c}{f\cs^2}\right)^{1/2}r,\label{eq:def_xi}
\end{equation}
and $\xi_c=\sqrt{26/3}$ \cite[]{TomidaPhD,2012PASJ...64..116K}\footnote{The total mass of this mass distribution diverges as $r\rightarrow \infty$, therefore this profile must be used within a finite radius.}. Then we need to estimate the initial central density of the cloud $\rho_c$ from the observed values.

For this purpose, we use two observed values: the mass of the central protostar $M_*$ and the specific angular momentum $j_e$ in the inner envelope where $j_e$ looks spatially constant. $M_*$ can be determined using the rotation profile in the Keplerian disk. In the inner envelope, we can safely assume that the enclosed mass within this region is dominated by the central protostar, i.e. $M_r\sim M_*$ where $j(r)\sim j_e$.

Because the specific angular momentum is almost conserved in the envelope, we can determine the initial location of the gas in the inner envelope using the specific angular momentum if we know its initial distribution. Observations indeed suggest that the distribution is more or less universal \citep{ohashi97,2013EAS....62...25B}:
\begin{equation}
 j_i (r)= 0.4\left(\frac{r}{1 [\rm pc]}\right)^{1.6} [\rm km\, pc\,s^{-1} ].\label{eq:j_c}
\end{equation}
From this equation, the initial radius of the gas element whose specific angular momentum is $j_{\rm e}$ is 
\begin{equation}
 r_{\rm e,i} = \left({\frac{j_{\rm e}}{0.4 {\rm [km \ pc\ s^{-1}]}}}\right)^{1/1.6} [\rm pc]\label{eq:r_ei}.
\end{equation}

By integrating Equation (\ref{eq:rho_r}), we obtain the enclosed mass within the radius $r_{e,i} = \xi_{\rm e,i}\cs\sqrt{f/(4\pi G \rho_{\rm c})}$, which is equal to $M_*$. The relation between $\xi_{\rm e,i}$ and $M_{\rm *}$ is given by the following equation:
\begin{equation}
  \frac{\xi_{\rm c}^3}{\xi_{\rm e,i}}\left[ \ln\left(\frac{\xi_{\rm e,i}}{\xi_{\rm c}} + \sqrt{1+
			  \frac{\xi_{\rm e,i}^2}{\xi_{\rm c} ^2}}\right) -
  \frac{\xi_{\rm e,i}}{\sqrt{\xi_{\rm e,i}^2+\xi_{\rm c}^2  }}\right] = \frac{GM_*}{f\cs^2 r_{\rm e,i}}.
\label{eq:xi}
\end{equation}
Solving Equation (\ref{eq:xi}) \footnote{Note that the Maximum value of the left hand side of Equation (\ref{eq:xi}) is about 2.5. 
If the protostar mass $M_{\rm *}$ is large and/or the specific angular momentum of the infalling envelope $j_{\rm e}$ is small ($r_{\rm e,i}$ is small), then we cannot solve Equation (\ref{eq:xi}).
In this case, the assumption of initial rotation profile $j_{\rm c}$ (Equation (\ref{eq:j_c})) or density enhanced parameter $f$ might be inappropriate. The simplest solution to this problem is increasing $f$.} for $\xi_{\rm e,i}$ and substituting it into Equation (\ref{eq:def_xi}), we obtain the initial central density,
\begin{equation}
 \rho_{\rm c} = \frac{f \xi_{\rm e,i}^2 \cs^2 }{4\pi G r^2_{\rm e,i}}.\label{eq:rho_c}
\end{equation}

\begin{figure}
\begin{center}
\scalebox{1}{\includegraphics{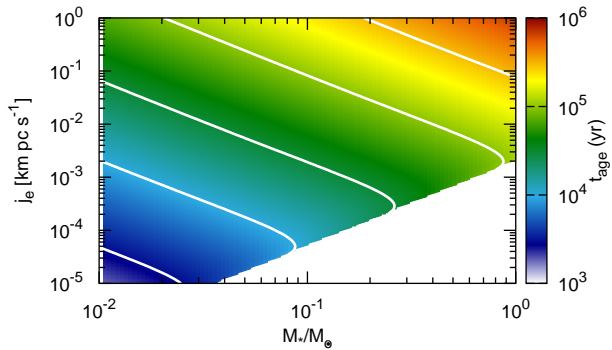}}
  \caption{Protostar age obtained from the analytic model { with $f=1.4$}. Horizontal axis is protostar mass and vertical axis is specific angular momentum in the constant specific angular momentum region. 
The solid white lines show $t_{\rm age}=3\times 10^3, \ 10^4, \ 3 \times 10^4, \ 10^5,\ 3\times 10^5$ yr from bottom to top.
In the white region, Equation (\ref{eq:xi}) has no real solution. 
}
\label{fig:tage}
\end{center}
\end{figure}

Thus all the parameters of the model are determined. Now we can apply our theoretical model proposed in Section~\ref{analytic_model} to estimate the ages of protostars. The current age of the system measured from the beginning of the collapse can be estimated by
\begin{equation}
 t_{\rm sys}= A(M_*)\sqrt{ \frac{r_{\rm e,i}^3}{2GM_*} }\int^1_0\frac{dx}{\sqrt{f^{-1}\ln x +
  x^{-1} -1} },
\label{eq:tsys}
\end{equation}
while the epoch of the formation of the protostar can be calculated by \begin{equation}
 t_{\rm pro}=  1.4\sqrt{\frac{3}{8\pi G \rho_c}}\int_0^1\frac{dx}
{\sqrt{f^{-1}\ln x +x^{-1} -1}}.
\label{eq:tpro}
\end{equation}
Then, the age of the protostar is $t_{\rm age} \equiv t_{\rm sys} -t_{\rm pro}$. 
Figure \ref{fig:tage} shows the protostar age obtained by this analytic model for various $M_{\rm *}$ and $j_{\rm e}$, where we adopt $f=1.4$.
The solid white lines show $t_{\rm age}=3\times 10^3, \ 10^4, \ 3 \times 10^4, \ 10^5,\ {\rm  and}\ 3\times 10^5$ yr from bottom to top.
The white region in Figure \ref{fig:tage} shows the region where we cannot solve Equation (\ref{eq:xi}).
The protostar age given in Figure \ref{fig:tage} is approximately fitted by
\begin{eqnarray}
 t_{\rm age} = 7.0\times 10^5 
  \left(\frac{M_*}{M_{\rm \odot}}\right)^{0.5}
  \left(\frac{j_e}{1\ [\rm km\ pc\ s^{-1}]}\right)^{0.31}\nonumber\\
 +7.2\times 10^{-9}\left(\frac{M_*}{M_{\rm \odot}}\right)^{8.4}
  \left(\frac{j_e}{1\ [\rm km\ pc\ s^{-1}]}\right)^{-4.6}
[\rm yr].
\end{eqnarray}
The difference of the estimated age obtained from this fitting formula and the age shown in Figure~\ref{fig:tage} is less than 10\%.

\subsection{L1527}
Here we take L1527 IRS as a well-studied example. 
L1527 is considered as a class 0 object.
Recent observations with ALMA \cite[]{2014ApJ...796..131O} indicated that the gas in the intermediate region ($10^2 \, {\rm AU}\ltsim r \ltsim 10^3\, {\rm AU}$) has a uniform specific angular momentum, $j_{\rm e}\sim 6.1 \times 10^{-4} \, {\rm pc \, km \, s^{-1}}$. 
 The observations showed that the rotational profile in the innermost region might indicate the existence of a circumstellar disk with a Keplerian rotation profile. In \cite{2014ApJ...796..131O}, they estimated the size of the disk is about (or smaller than) $60\, {\rm AU}$ and the mass of the central object is about $M_{*}\sim 0.3 M_\odot$. 
We also assume that the gas temperature of the cloud is $10\, {\rm K}$.
Applying our model to this observation, the gas elements in the intermediate region originally come from $r_{\rm e,i} \sim 3.6\times 10^3 \, {\rm AU}$. 
From Equations (\ref{eq:xi}) and (\ref{eq:rho_c}), the central density of the initial molecular cloud is $\rho_{\rm c} \sim 1.8\times 10^{-18} \,  {\rm g\, cm^{-3}}$.
We obtain $t_{\rm pro}\sim 1.1\times 10^5$ yr, $t_{\rm sys} \sim 1.5\times 10^5 $ yr and the protostar age $t_{\rm age} \sim 3.7\times 10^4$ yr from Equations (\ref{eq:tsys}) and (\ref{eq:tpro}).
This result implies that this object is still very young, which is consistent with observational facts, especially its small circumstellar disk.

Note that, although this estimate is not very sensitive to the value of $f$ unless $f$ is very close to unity, there still is an uncertainty of a factor of 2 (see Section \ref{discussion}).  
These observations still have significant uncertainty even with ALMA, because the base lines available in Cycle-0 observations were limited and therefore the observations suffer from significant missing fluxes in the large scale ($r > 500 \, {\rm AU}$). The argument presented in this section should be considered as a demonstration how to use our model to estimate ages of protostars. More systematic and high-quality observations are highly demanded, and our model will be useful to study the dynamics of protostar formation by comparing the model with data obtained from such future observations.

\subsection{TMC-1A (L1534)}
Another well-studied is TMC-1A. This object is more evolved than L1527 IRS, and is classified as a Class I object \cite[cf.][]{2013ApJ...772...22Y}.
The rotation velocity distribution of TMC-1A is given by \citet{2015arXiv150807013A}.
The radius of the Keplerian disk is about 100 AU and the central star mass is estimated as 0.68$M_{\rm \odot}$.
The specific angular momentum of the inner edge of the infall envelope (or outer edge of the disk) is $1.2\times 10^{-3} \ {\rm km \ pc\ s^{-1}}$.
From Equation (\ref{eq:r_ei}), the initial radius of the envelope is
$\sim 5.5\times 10^3$ AU.
From Equations (\ref{eq:xi}) and (\ref{eq:rho_c}), the initial central density of the cloud is $\rho_{\rm c} \sim  2.4\times 10^{-18} {\rm g\ cm^{-3}}$.
As a result, estimated ages are $t_{\rm pro}\sim 9.8\times 10^4$ yr, $t_{\rm sys} \sim 1.8\times 10^5 $ yr and $t_{\rm age} \sim 7.8\times 10^4$ yr from Equations (\ref{eq:tsys}) and (\ref{eq:tpro}).

\subsection{B335}
Recently \cite{2015ApJ...812..129Y} reported non-detection of circumstellar disks down to
~10 AU scale for B335. They suggested that this object is extremely young and/or strongly
affected by magnetic fields.
The observed specific angular momentum is $\sim 4.3\times 10^{-5} \ {\rm km\ pc\ s^{-1}}$ at 100 AU. 
Similar specific angular momentum $\sim (3 \-- 5) \times 10^{-5} \ {\rm km\ pc \ s^{-1}}$ is obtained at 10 AU.
Thus, we adopt the specific angular momentum $\sim 4.3\times 10^{-5} \ {\rm km\ pc \ s^{-1}}$ as a constant specific angular momentum of the envelope.
The central star mass is $\sim 0.05 M_{\rm \odot}$.
The large-scale rotation velocity of B335 is obtained from single-dish
 observations \cite[]{1999ApJ...518..334S,2011ApJ...742...57Y,2013ApJ...765...85K}.
Based on these observations, the initial rotational velocity of B335 is assumed to be a rigid-body rotation with an angular velocity of $0.8\  {\rm km \ s^{-1} \ pc^{-1}}$.
From Equation (\ref{eq:r_ei}), the initial radius of the envelope is
$\sim 1.5\times 10^3$ AU.
From Equations (\ref{eq:xi}) and (\ref{eq:rho_c}), the initial central density of the cloud is $\rho_{\rm c} \sim  2.5\times 10^{-18} {\rm g\ cm^{-3}}$.
Thus,  Equations (\ref{eq:tsys}) and (\ref{eq:tpro}) yield $t_{\rm pro}\sim 9.6\times 10^4$ yr, $t_{\rm sys} \sim 1.1\times 10^5 $ yr and $t_{\rm age} \sim 9.6\times 10^3$ yr.
Thus, our model suggests that this object is indeed extremely young, although we cannot account for the effects of magnetic fields.

\subsection{Comparison with the Outflow Time Scale}
For young Class-0 objects, the molecular outflows are often used to estimate the ages, simply dividing the spatial extent by the characteristic speed of the outflows. 
The maximum velocity or the average velocity are often used in the literature for the characteristic speed, but it is not trivial to determine which one is more appropriate for estimating the age of the system. When the maximum velocity is considered, this results in shorter dynamical time-scales.

For L1527 IRS, \cite{nar12} derived the dynamical time scale of the outflow based on the average velocity, which is about $3 \times 10^4 \, {\rm yr}$. This is in a good agreement with our estimate. 

\cite{1996ApJ...471..308C} estimated the dyanmical time scale of TMC-1A to be 3,700 years, while it is estimated to be $1-2 \times 10^4 \, {\rm yr}$ using the average velocity \citep{nar12}. These are both much younger than our estimate, but it is possibly because the outflow is so faint that only a portion of that is detected, or because the outflow becomes not prominent as it interacts with the ISM. In either way, the age estimate based on the outflow should be considered as a lower limit.

For B335, \cite{1988ApJ...327L..69H} \cite[see also][]{2015ApJ...812..129Y} estimated its age to be about $1.3 - 3.3 \times 10^4 \, {\rm yr}$ using the average velocity of $\sim 13 \, {\rm km \, s^{-1}}$. Although the age is much longer than our estimate, indeed very fast components up to $\sim 160 {\rm km \, s^{-1}}$ are detected toward this object \cite[]{2010ApJ...710.1786Y}. Therefore we suspect that the use of the average velocity results in overestimating the age of this object. Also it should be noted that this object is close to edge-on configuration and therefore the inclination error can significantly affect the velocity. Another possible interpretation is that this system is strongly magnetized and angular momentum is removed very efficiently. In this case, our model should result in underestimating the age.

\section{Discussion}
\label{discussion}

\subsection{Dependence on $f$}
\begin{figure}
\begin{center}
\scalebox{1}{\includegraphics{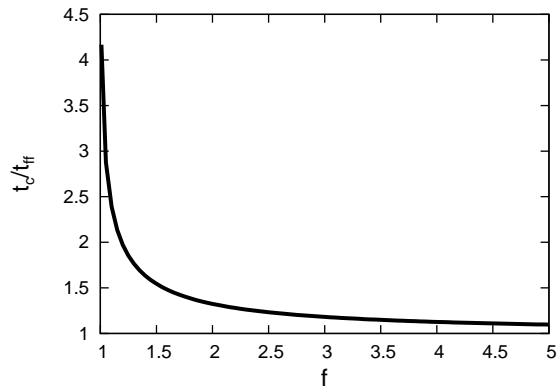}}
\caption{
Ratio of the infall time to free-fall time as a function of $f$ given by Equation (\ref{tc}).
}
\label{alpha_dependence}
\end{center}
\end{figure}

In this work, we introduce $f$ as a parameter of the gravitational instability of initial molecular clouds.
This parameter gives the ratio between the pressure force and gravitational force (Equation (\ref{eq:dvrdt})).
Although the infall time depends on $f$ (see Equation (\ref{pos})),we assume $f =1.4 $ for comparison with the observations.
Figure \ref{alpha_dependence} shows the ratio of the infall time to free-fall time as a function of $f$ given by Equation (\ref{tc}).
This ratio is much larger than unity around $f \sim 1$
and asymptotically converges to unity as $f$ increases.
The ratio is smaller than 2 for $f \gtrsim 1.2$.
The uncertainty associated with the parameter $f$, hence, is smaller than a factor of 2 as long as the realistic $f$ is not very close to unity.

Observations of molecular clouds suggest that $f$ is higher than unity.
\citet{2005MNRAS.360.1506K} have fitted the observed density profiles of molecular clouds by BE sphere and found that the effective temperature that gives the best fit of the density profile is different from the observed temperature.
The ratio of these two temperatures, which indicates the relation between the actual gravitational force and thermal pressure, can be used to infer $f$.

The density profile of the BE sphere is almost constant inside the radius $r_{\rm c}$,
\begin{equation}
 r_{\rm c} = \sqrt{\frac{\cs^2}{4\pi G \rho_{\rm c}}},
\label{eq:r_c}
\end{equation}
where $\rho_{\rm c}$ is the central density.
The radius $r_{\rm c}$ increases as the temperature increases.
For the clouds observed by \citet{2005MNRAS.360.1506K}, $r_{\rm c}$ is too small when we use the temperature obtained from the observation, $T_{\rm obs} \sim 10$ K. The effective temperature $T_{\rm eff}$ exceeds $T_{\rm obs}$ by a factor of $\sim 2 \-- 8$.
On the other hand, with the density distribution we adopt in this work, density is almost constant inside the radius $\sqrt{\cs^2f/4\pi G \rho_{\rm c}}$. Thus, we can fit the observed density profile when we adopt $T=T_{\rm obs} \sim 10$ K and $f >1$.
The relation of $f$, $T_{\rm obs}$ and $T_{\rm eff}$ is
$f = T_{\rm eff}/T_{\rm obs}$.
Since \citet{2005MNRAS.360.1506K} have obtained $2\lesssim T_{\rm eff}/{T_{\rm obs}} \lesssim 8$,
observations of the cloud cores suggest that the parameter $f$ is typically higher than unity and the uncertainty associated with $f$ is smaller than two.

\subsection{Effect of the external region of BE sphere}
As described above, observations of molecular cloud cores indicate that the density profile is well fitted by that of the BE spheres \cite[e.g.][]{2001Natur.409..159A,2005MNRAS.360.1506K}.
Thus, we develop the analytic model for the gravitational collapse of a molecular cloud core whose density profile is similar to that of BE sphere.
In general, star forming cloud cores are not isolated but embedded in molecular clouds.
Recent observations have revealed that the cloud cores are in the filamentary clouds \cite[]{2010A&A...518L.102A}.
In this work, we do not take into account the effect of the external region of molecular cloud cores on the gravitational collapse.
Because the timescale of the gravitational collapse at the center of the molecular cloud is shorter than the timescale at the outer region, the external region is not expected to affect the collapse at the center in the early phase of star formation.
For example, the model with $f=1.4$ used in Section \ref{analytic_model} has the free-fall time of $5.4\times 10^4$ yr at the center.
On the other hand, at 10000 au which is the typical size of a molecular cloud core, it is $1.5\times 10^5$ yr.
Therefore, the external region is not expected to affect the gravitational collapse and the structure of the envelope when the protostar age is smaller than about $10^5$ yr.

\cite{2003ApJ...592..188B} and \cite{2004IAUS..221..201H} have pointed out that even when the observed column density profile of the cloud core is well fitted by that of the BE sphere, the actual density distribution of the core can be different from that of the BE sphere.
When the initial condition significantly deviates from the BE sphere adopted here (for example, when the gravitational collapse of the core starts because of cloud-cloud collision or when the cloud core contains strong turbulence), our simple analytic model cannot be applied.
Also, when the initial angular momentum distribution has a discontinuity (e.g., when gas with a completely different angular momentum flows into the cloud core), such a discontinuous structure will remain in the infalling envelope because it cannot be smoothed out by elongation.
The investigation of such a complex structure of infalling envelope is beyond the scope of this study.

\subsection{Initial rotation velocity}
There is another uncertainty associated with the initial specific angular momentum distribution.
In Section \ref{comparison}, we use the relation between the radius of the molecular clouds and their rotation velocity obtained from the observations of many NH$_3$ cores \cite[]{ohashi97} as the initial rotation profile (Equation (\ref{eq:j_c})).
This is not the spatially-resolved rotation velocity distribution of a single molecular cloud.
L1544 is one of the molecular clouds for which the spatially resolved rotation velocity profile has been obtained from observations \cite[]{1999ApJ...518L..41O}.
The specific angular momentum of L1544 is fitted by rigid rotation as follows:
\begin{equation}
 j_{\rm c} \sim 1.2 \left(\frac{r}{1\ [\rm pc]}\right)^2 [\rm km \ pc \ s^{-1}]\label{eq:rot_l1544}
\end{equation}
When we adopt this as the initial specific angular momentum distribution  of L1527, the estimated age is $t_{\rm age} \sim 4.2 \times 10^4$ yr. This age is similar to the age obtained in Section \ref{comparison} ($3.7\times 10^4$ yr).
When we apply Equation (\ref{eq:rot_l1544}) as the initial specific angular distribution of TMC-1A, we obtain $t_{\rm age} \sim 7.9 \times 10^4$ yr. This age is also similar to the age obtained in Section \ref{comparison} ($7.8 \times 10^4$ yr).

It is technically possible to use the observed large-scale velocity gradient of each molecular cloud cores as the initial rotation profile.
For example, the large-scale ($\sim 0.1$ pc) velocity gradient of L1527 obtained from ${\rm N_2H^+}$ observation with the IRAM 30m single-dish telescope by \citet{tobin11} 
is about $2.2 \,{\rm km \, s^{-1} \, pc^{-1}}$. 
However, interestingly enough, this large-scale velocity gradient is often mis-aligned or even counter-rotating to the rotation in the smaller scale, probably reflecting the nature of turbulence in the interstellar media \cite[]{lrs81}. 
Nevertheless, we can consider this as the characteristic specific angular momentum distribution of L1527. 
In this case, the protostar age of L1527 is $t_{\rm age}\sim 3.7\times 10^4$ yr.
This is in good agreement with the age obtained in Section \ref{comparison}.
The age of L1527 estimated in this work and the dynamical age obtained in \cite{nar12} is summarized in Table \ref{AgesL1527}.
The results are apparently not very sensitive to the initial rotation in each object, because the initial rotation is more or less universal.
\begin{table*}
\begin{center}
\caption[]{
Protostar ages of L1527 estimated in this work and the dynamical age obtained in \cite{nar12}.
}
\label{AgesL1527}
 \begin{tabular}{lcc}\hline\hline
  && Age  [yr]\\ \hline
  Initial specific angular momentum $[\rm km\ pc\ s^{-1}]$ & 
      $0.4\left(\frac{r}{1[\rm pc]}\right)^{1.6}$& $3.7\times 10^4$\\ 
  & $1.2\left(\frac{r}{1[\rm pc]}\right)^2$& $4.2\times 10^4$\\
  &$2.2\left(\frac{r}{1[\rm pc]}\right)^2$& $3.7\times 10^4$\\
\hline
  Dynamical age \cite[]{nar12}&& $3\times 10^4$\\ \hline
 \end{tabular}
\end{center}
\end{table*}

\subsection{Effect of magnetic field}
In this work, we neglect effects of magnetic field for simplicity.
The effect of the magnetic field on the star formation is actively investigated using numerical simulations \cite[e.g.][]{2011PASJ...63..555M, 2014MNRAS.438.2278M, 2011ApJ...738..180L,2015ApJ...810L..26T,2016MNRAS.457.1037W}.
When the magnetic field in clouds is strong enough, there are two important effects of the magnetic field on the infalling envelope.
One is that the infall velocity decreases due to magnetic force.
The infall velocity of L1527 estimated from the observation is smaller than free-fall velocity estimated from the central star mass \cite[]{2014ApJ...796..131O}.
This difference may be caused by the magnetic fields.
When the infall velocity decreases due to the magnetic pressure, our analytic model underestimates $t_{\rm sys}$ and $t_{\rm pro}$.
The other is the angular momentum transfer of the infalling envelope by magnetic braking.
The angular momentum of the infalling envelope in the midplane is transferred to the upper region due to the magnetic braking.
As a result, the observed angular momentum is smaller than the initial angular momentum, and we underestimate the initial radius of the infalling envelope if we assume the conservation of the angular momentum in the infalling envelope. 
Thus, our analytic model should give the lower limit of $t_{\rm sys}$ and $t_{\rm pro}$.

To estimate the protostar age in the strongly magnetized clouds, we need to extend our model to take into account the effect of the magnetic field.

\section{Conclusions}
We propose a simple analytic model to describe the structure and evolution of collapsing clouds. Although this model is so simple that it considers only gravity and gas pressure with an approximated treatment in quasi spherically symmetric configuration, it can successfully reproduce the result of the hydrodynamic simulation and provide a reasonable fit to observational data. While this model includes neither the effects of rotation nor magnetic fields, it is still applicable to the outer region where the rotation velocity is still small and magnetic braking is not significant.
We showed that the region with constant specific angular momentum can be formed as a consequence of strong elongation in a run-away collapse of a cloud, regardless of the initial angular momentum distribution.
We have also provided a simple method to estimate the ages of protostars based on our model, using information of the rotational profile only. This information can be obtained from easily accessible molecular line observations.
We applied our model to L1527, TMC-1A and B335, and obtained the protostar ages of, $3.7\times 10^4$ yr, $7.8\times 10^4$ yr, and $9.6\times 10^3$ yr, respectively.
Our model will be useful to study the dynamics in the infalling envelope of young stellar objects and related circumstellar disk formation, especially for the present and future observation with ALMA.

\section*{Acknowledgments}
We are grateful to the anonymous referee for useful comments which helped improve the manuscript.
TZS and KT are supported by Japan Society for the Promotion of Science (JSPS) Research Fellowship for Young Scientists. The numerical simulation in \S3 was performed using NEC SX-9 at the Center for Computational Astrophysics of National Astronomical Observatory of Japan.
This research also used computational resources of the HPCI system provided by the Cyberscience Center, Tohoku University, the Cybermedia Center, Osaka University, the Earth simulator, JAMSTEC through the HPCI System Research Project
(Project ID:hp160079).

\end{document}